# A PROBLEM OF HYPOTHETICAL EMERGING OF COSMIC BACKGROUND RADIATION PHOTONS ON HORIZON IN THE STANDARD MODEL OF UNIVERSE

## Vladimír Skalský

Faculty of Materials Science and Technology of the Slovak Technical University, 917 24 Trnava, Slovakia, skalsky@mtf.stuba.sk

**Abstract.** The present temperature of cosmic background radiation and the present number density of photons of cosmic background radiation in the observed expansive and isotropic relativistic Universe is in the standard model of universe explained by the assumption of emergence of the photons of cosmic background radiation on the horizon (of the most remote visibility). However, the physical analysis shows unambiguously that this assumption contradicts the special theory of relativity and the quantum mechanics.

According to the present cosmological literature, the matter and the radiation were separated at
*the cosmological time of the Universe at the end of radiation (photon) era*

$$t_{end} \sim 3 \times 10^5 \text{ yr} \tag{1}$$

at
*the temperature of the Universe at the end of radiation era*

$$T_{end} \sim 3 \times 10^3 \text{ K} . \tag{2}$$

To the relation (1) correspond
*the gauge factor of the Universe at the end of radiation era*

$$a_{end} \sim 3 \times 10^{21} \text{ m} \tag{3}$$

and
*the volume of the Universe at the end of radiation era*

$$V_{end} \sim 10^{65} \text{ m}^3 . \tag{4}$$

The number of photons of the absolutely black body radiation (Landau and Lifshitz 1964)

$$n_\gamma = \frac{V}{\pi^2 c^3} \int_0^\infty \frac{\omega^2 \, d\omega}{e^{\hbar\omega/kT}-1} = \frac{V k^3 T^3}{\pi^2 c^3 \hbar^3} \int_0^\infty \frac{x^2 \, dx}{e^x - 1}, \tag{5}$$

where $V$ is the volume, $k$ is the Boltzmann constant and $T$ is the temperature of absolutely black body.

From the relations (2) and (5) it results
*the number density of photons at the end of radiation era*

$$n_{\gamma-end} \sim 6 \times 10^{17} \text{ m}^{-3} . \tag{6}$$

From the relations (4) and (6) it results:
*the total number of photons at the end of radiation era*

$$Ó_{\gamma-end} \sim 6 \times 10^{82} . \tag{7}$$

According to the present observations,
*the present temperature of cosmic background radiation* (Mather *et al.* 1994)

$$T_{CBR-pres} = 2.726 \pm 0.010 \text{ K} . \tag{8}$$

From the relations (5) and (8) it results
*the present number density of photons of cosmic background radiation*

$$n_{\gamma-pres} = 4.1096 \times 10^8 \text{ m}^{-3} . \tag{9}$$

*Note*: The Particle Data Group (1996) give

$$n_{\gamma-pres} = 4.1089 \times 10^8 \text{ m}^{-3} . \tag{9x}$$

At present time is estimated

*the present gauge factor of Universe*

$$a_{pres} \sim 10^{26} \text{ m}. \tag{10}$$

To the relation (10) corresponds

*the present volume of Universe*

$$V_{pres} \sim 4 \times 10^{78} \text{ m}^3. \tag{11}$$

From the relations (9), or (9x), and (11) it results

*the present total number of photons of cosmic background radiation*

$$Ó_{\gamma-pres} \sim 2 \times 10^{87}. \tag{12}$$

From comparison of the relations (7) and (12) it results that the total photons number of cosmic background radiation in the matter era increased. The ratio of the total number of cosmic background radiation photons at the present time and of those at the end of radiation era is

$$\frac{Ó_{\gamma-pres}}{Ó_{\gamma-end}} \sim \frac{2 \times 10^{87}}{6 \times 10^{82}} \sim 3 \times 10^4. \tag{13}$$

There exist two basic approaches to the attempt of explaining the increase in number of photons of cosmic background radiation:

A. *Extensive*, according to which the increase in number of photons of cosmic background radiation is caused by *the emergence of the photons of cosmic background radiation on the horizon (of the most remote visibility), (the horizon of events)*.

B. *Intensive*, according to which the increase in number of photons of cosmic background radiation is the result of *the permanent constant maximum possible creation of matter (and radiation)* (Skalský 1994, 1997).

In *the standard model of universe* the increase in number of photons of cosmic background radiation is explained by the extensive assumption A. Therefore – under this assumption – from the relation (13) it results that since the beginning of matter era approximately 30 000-multiple of the photons of cosmic background radiation visible at the time $t_{end}$ (1) has emerged on the horizon (of the most remote visibility).

According to the standard model of universe, the Universe at the distance of gauge factor $a$ expands at the velocity

$$v \sim c. \tag{14}$$

Therefore, according to the extensive assumption A, the photons of cosmic background radiation emerging on the horizon (of the most remote visibility) have been radiated by the sources which at the time $t_{end}$ (1) expand relatively to our location at the radial velocities

$$v \geq c. \tag{15}$$

For the Doppler red shift $z_D$ of photons radiated from the source expanding at the velocity $v$ the well-known relations (Einstein 1905) are valid:

$$z_D = \sqrt{\frac{c+v}{c-v}} - 1 \equiv \frac{1+\frac{v}{c}}{\sqrt{1-\frac{v^2}{c^2}}} - 1. \tag{16}$$

The Doppler red shift $z_D$ in the relations (16) can obtain the values in the interval

$$0 < z_D < \infty, \tag{17}$$

i.e. it is defined for the photons radiated by the sources moving at the radial velocity $v$ in the interval

$$0 < v < c. \tag{18}$$

Therefore, according to the relations (16), any considerations on the photons of cosmic background radiation coming from the sources behind the horizon (of the most remote visibility) – which at the time separation matter and radiation $t_{end}$ (1) moved relatively to our location at the radial velocities $v$ (15) – have no physical sense.

**Conclusions**

The present temperature of cosmic background radiation $T_{CBR-pres}$ (8) and the present number density of photons of cosmic background radiation $n_{\gamma-pres}$ (9) in our observed expansive and isotropic relativistic Universe principally cannot be explained by the extensive assumption of emergence of the photons of cosmic background radiation on the horizon (of the most remote visibility) because this assumption contradicts the special theory of relativity and the quantum mechanics.